**#275**

**Database Benchmarks**

Jérôme Darmont[*]

ERIC, University of Lyon 2, France

5 avenue Pierre Mendès-France

69676 Bron Cedex

France

[jerome.darmont@univ-lyon2.fr](jerome.darmont@univ-lyon2.fr)

Phone: +33 478 774 403

Fax: +33 478 772 375

INTRODUCTION

Performance measurement tools are very important, both for designers and users of Database Management Systems (DBMSs). Performance evaluation is useful to designers to determine elements of architecture, and more generally to validate or refute hypotheses regarding the actual behavior of a DBMS. Thus, performance evaluation is an essential component in the development process of well-designed and efficient systems. Users may also employ performance evaluation, either to compare the efficiency of different technologies before selecting a DBMS, or to tune a system.

Performance evaluation by experimentation on a real system is generally referred to as benchmarking. It consists in performing a series of tests on a given DBMS to estimate its performance in a given setting. Typically, a benchmark is constituted of two main elements: a database model (conceptual schema and extension) and a workload model (set of read and write operations) to apply on this database, following a predefined protocol. Most benchmarks also include a set of simple or composite performance metrics such as response time, throughput, number of input/output, disk or memory usage, etc.

The aim of this article is to present an overview of the major families of state-of-the-art database benchmarks, namely: relational benchmarks, object and object-relational benchmarks, XML benchmarks, and decision-support benchmarks; and to discuss the issues, tradeoffs and future trends in database benchmarking. We particularly focus on XML and decision-support benchmarks, which are currently the most innovative tools that are developed in this area.

BACKGROUND

Relational benchmarks

In the world of relational DBMS benchmarking, the Transaction Processing Performance Council (TPC) plays a preponderant role. The mission of this non-profit organization is to issue standard benchmarks, to verify their correct application by users, and to regularly publish performance tests results. Its benchmarks all share variants of a classical business database (*customer-order-product-supplier*) and are only parameterized by a scale factor that determines the database size (*e.g.*, from 1 to 100,000 GB).

The TPC benchmark for transactional databases, TPC-C (TPC, 2005a), has been in use since 1992. It is specifically dedicated to On-Line Transactional Processing (OLTP) applications, and features a complex database (nine types of tables bearing various structures and sizes), and a workload of diversely complex transactions that are executed concurrently. The metric in TPC-C is throughput, in terms of transactions.

There are currently few credible alternatives to TPC-C. Although, we can cite the Open Source Database Benchmark (OSDB), which is the result of a project from the free software community (SourceForge, 2005). OSDB extends and clarifies the specifications of an older benchmark, AS$^3$AP. It is available as free C source code, which helps eliminating any ambiguity relative to the use of natural language in the specifications. However, it is still an ongoing project and the benchmark's documentation is very basic. AS$^3$AP's database is simple: it is composed of four relations whose size may vary from 1 GB to 100 GB. The workload is made of various queries that are executed concurrently. OSDB's metrics are response time and throughput.

Object-oriented and object-relational benchmarks

There is no standard benchmark for object-oriented DBMSs. However, the most frequently cited and used, OO1 (Cattel, 1991), HyperModel (Anderson *et al.*, 1990), and chiefly OO7 (Carey et al, 1993), are *de facto* standards. These benchmarks mainly focus on engineering applications (*e.g.*, computer-aided design, software engineering). They range from OO1, which bears a very simple schema (two classes) and only three operations, to OO7, which is more generic and proposes a complex and tunable schema (ten classes), as well as fifteen complex operations. However, even OO7, the more elaborate of these benchmarks, is not generic enough to model other types of applications, such as financial, multimedia or telecommunication applications (Tiwary *et al.*, 1995). Furthermore, its complexity makes it hard to understand and implement. To circumvent these limitations, the OCB benchmark has been proposed (Darmont & Schneider, 2000). Wholly tunable, this tool aims at being truly generic. Still, the benchmark's code is short, reasonably easy to implement, and easily portable. Finally, OCB has been extended into the Dynamic Evaluation Framework (DEF), which introduces a dynamic component in the workload, by simulating access pattern changes using configurable styles of changes (He & Darmont, 2005).

Object-relational benchmarks such as BUCKY (Carey *et al.*, 1997) and BORD (Lee *et al.*, 2000) are query-oriented and solely dedicated to object-relational systems. For instance, BUCKY only proposes operations that are specific to these systems, considering that typical object navigation is already addressed by object-oriented benchmarks. Hence, these benchmarks focus on queries implying object identifiers, inheritance, joins, class and object references, multivalued attributes, query unnesting, object methods, and abstract data types.

XML benchmarks

Since there is no standard model, the storage solutions for XML (eXtended Markup Language) documents that have been developed since the late nineties bear significant differences, both at the conceptual and the functionality levels. The need to compare these solutions, especially in terms of performance, has lead to the design of several benchmarks with diverse objectives.

X-Mach1 (Böhme & Rahm, 2001), XMark (Schmidt *et al.*, 2002), XOO7 (an extension of OO7; Bressan *et al.*, 2002) and XBench (Yao *et al.*, 2004) are so-called application benchmarks. Their objective is to evaluate the global performances of an XML DBMS, and more particularly of its query processor. Each of them implements a mixed XML database that is both data-oriented (structured data) and document-oriented (in general, random texts built from a dictionary). However, except for XBench that proposes a true mixed database, their orientation is more particularly focused on data (XMark, XOO7) or documents (X-Mach1). These benchmarks also differ in:

- the fixed or flexible nature of the XML schema (one or several Document Type Definitions or XML schemas);
- the number of XML documents used to model the database at the physical level (one or several);
- the inclusion or not of update operations in the workload.

We can also underline that only XBench helps in evaluating all the functionalities offered by the XQuery language.

Micro-benchmarks have also been proposed to evaluate the individual performances of basic operations such as projections, selections, joins, and aggregations, rather than more complex queries. The Michigan Benchmark (Runapongsa *et al.*, 2002) and MemBeR (Afanasiev *et al.*, 2005) are made for XML documents storage solution designers, who can isolate critical issues to optimize, rather than for users seeking to compare different systems. Furthermore, MemBeR proposes a methodology for building micro-databases, to help users in adding datasets and specific queries to a given performance evaluation task.

Decision-support benchmarks

Since decision-support benchmarks are currently a *de facto* subclass of relational benchmarks, the TPC again plays a central role in their standardization. TPC-H (TPC, 2005c) is currently their only decision-support benchmark. It exploits a classical *product-order-supplier* database schema; as well as a workload that is constituted of twenty-two SQL-92, parameterized, decision-support queries and two refreshing functions that insert tuples into and delete tuples from the database. Query parameters are randomly instantiated following a uniform law. Three primary metrics are used in TPC-H. They describe performance in terms of power, throughput, and a combination of these two criteria.

Data warehouses nowadays constitute a key decision-support technology. However, TPC-H's database schema is not a star-like schema that is typical in data warehouses. Furthermore, its workload does not include any On-Line Analytical Processing (OLAP) query. TPC-DS, which is currently under development (TPC, 2005b), fills in this gap. Its schema represents the decision-support functions of a retailer under the form of a constellation schema with several fact tables and shared dimensions. TPC-DS' workload is constituted of four classes of queries: reporting queries, *ad-hoc* decision-support queries, interactive OLAP queries, and extraction queries. SQL-99 query templates help in randomly generating a set of about five hundred queries, following non-uniform distributions. The warehouse maintenance process includes a full ETL (Extract, Transform, Load) phase, and handles dimensions according to their nature (non-static dimensions scale up while static dimensions are updated). One primary throughput metric is proposed in TPC-DS. It takes both query execution and the maintenance phase into account.

As in all the other TPC benchmarks, scaling in TPC-H and TPC-DS is achieved through a scale factor that helps defining the database's size (from 1 GB to 100 TB). Both the database schema and the workload are fixed.

There are, again, few decision-support benchmarks out of the TPC, and their specifications are rarely integrally published. Some are nonetheless of interesting. APB-1 is presumably the most famous. Published by the OLAP council, a now inactive organization founded by OLAP vendors, APB-1 has been intensively used in the late nineties. Its warehouse dimensional schema is structured around four dimensions: *Customer*, *Product*, *Channel*, and *Time*. Its workload of ten queries is aimed at sale forecasting. APB-1 is quite simple and proved limited to evaluate the specificities of various activities and functions (Thomsen, 1998). It is now difficult to find.

Eventually, while the TPC standard benchmarks are invaluable to users for comparing the performances of different systems, they are less useful to system engineers for testing the effect of various design choices. They are indeed not tunable enough and fail to model different data warehouse schemas. By contrast, the Data Warehouse Engineering Benchmark (DWEB) helps in generating various *ad-hoc* synthetic data warehouses (modeled as star, snowflake, or constellation schemas) and workloads that include typical OLAP queries (Darmont *et al.*, 2005a). DWEB is fully parameterized to fulfill data warehouse design needs.

ISSUES AND TRADEOFFS IN DATABASE BENCHMARKING

Gray (1993) defines four primary criteria to specify a "good" benchmark:

1. *relevance:* the benchmark must deal with aspects of performance that appeal to the largest number of potential users;
2. *portability:* the benchmark must be reusable to test the performances of different DBMSs;
3. *simplicity:* the benchmark must be feasible and must not require too many resources;
4. *scalability:* the benchmark must adapt to small or large computer architectures.

In their majority, existing benchmarks aim at comparing the performances of different systems in given experimental conditions. This helps vendors in positioning their products relatively to their competitors', and users in achieving strategic and costly software choices based on objective information. These benchmarks invariably present fixed database schemas and workloads. Gray's scalability factor is achieved through a reduced number of parameters that mainly allow varying the database size in predetermined proportions. It is notably the case of the unique scale factor parameter that is used in all the TPC benchmarks.

This solution is simple (still according to Gray's criteria), but the relevance of such benchmarks is inevitably reduced to the test cases that are explicitly modeled. For instance, the typical *customer-order-product-supplier* that is adopted by the TPC is often unsuitable to appli-cation domains other than management. This leads benchmark users to design more or less elaborate variants of standard tools, when they feel these are not generic enough to fulfill particular needs. Such users are generally not confronted to software choices, but are rather designers who have quite different needs. They mainly seek to evaluate the impact of architectural choices, or performance optimization techniques, within a given system or a family of systems. In this context, it is essential to multiply experiments and test cases, and a monolithic benchmark is of reduced relevance.

To enhance the relevance of benchmarks aimed at system designers, we propose to extend Gray's scalability criterion to *adaptability*. A performance evaluation tool must then be able to propose various database or workload configurations, to allow experiments to be performed in various conditions. Such a tool may be qualified as a benchmark generator, or as a tunable or generic benchmark. However, aiming at a better adaptability is mechanically detrimental to a benchmark's simplicity. This criterion though remains very important, and must not be neglected when designing a generic tool. It is thus necessary to devise means of achieving a good adaptability, without sacrificing simplicity too much. In summary, a satisfying tradeoff must be reached between these two orthogonal criteria.

We have been developing benchmarks following this philosophy for almost ten years. The first one, the Object Clustering Benchmark (OCB), was originally designed to evaluate the performances of clustering algorithms within object-oriented DBMSs. By extending its clustering-oriented workload, we made it generic. Furthermore, its database and workload are wholly tunable, through a collection of comprehensive but easily set parameters. Hence, OCB can be used to model many kinds of object-oriented database applications. In particular, it can simulate the behavior of the other object-oriented benchmarks.

Our second benchmark is the DWEB data warehouse benchmark. DWEB's parameters help users in selecting the data warehouse architecture and workload they need in a given context. To solve the adaptability *vs.* simplicity dilemma, we divided the parameter set into two subsets. Low-level parameters allow an advanced user to control everything about data warehouse generation. However, their number can increase dramatically when the schema gets larger. Thus, we designed a layer of high-level parameters that may be easily understood and set up, and that are in reduced number. More precisely, these high-level parameters are average values for the low-level parameters. At database generation time, the high-level parameters are automatically exploited by random functions to set up the low-level parameters.

FUTURE TRENDS

The development of XML-native DBMSs is quite recent, and a tremendous amount of research is currently in progress to help them becoming a credible alternative to XML-compatible, relational DMBSs. Several performance evaluation tools have been proposed to support this effort. However, research in this area is very dynamic, and new benchmarks will be needed to assess the performance of the latest discoveries. For instance, Active XML incorporates web services for data integration (Abiteboul *et al.*, 2002). An adaptation of existing XML benchmarks that would exploit the concepts developed in TPC-App, could help in evaluating the performance of an Active XML platform.

No XML benchmark is currently dedicated to decision-support either, while many XML data warehouse architectures have been proposed in the literature. We are currently working on a benchmark called XWB, which is aimed at evaluating the performances of such research pro-posals. Furthermore, there is a growing need in many decision-support applications (*e.g.*, customer relationship management, marketing, competition monitoring, medicine) to exploit complex data, *i.e.*, in summary, data that are not only numerical or symbolic. XML is particularly adapted to describe and store complex data (Darmont *et al.*, 2005b) and further adaptations of XML decision-support benchmarks would be needed to take them into account.

Finally, a lot of research also aims at enhancing the XQuery language, for instance with update capabilities, or with OLAP operators. Existing XML and/or decision-support benchmarks will also have to be adapted to take these new features into account.

CONCLUSION

Benchmarking is a small field, but it is nonetheless essential to database research and industry. It serves both engineering or research purposes, when designing systems or validating solutions; and marketing purposes, when monitoring competition and comparing commercial products.

Benchmarks might be subdivided in three classes. First, standard, *general-purpose benchmarks* such as the TPC's do an excellent job in evaluating the global performance of systems. They are well-suited to software selection by users and marketing battles by vendors, who try to demonstrate the superiority of their product at one moment in time. However, their relevance drops for some particular applications that exploit database models or workloads that are radically different from the ones they implement. *Ad-hoc benchmarks* are a solution. They are either adaptations of general-purpose benchmarks, or specifically designed benchmarks such as the XML micro-benchmarks we described above. Designing myriads of narrow-band benchmarks is not time-efficient, though; and trust in yet another new benchmark might prove limited in the database community. Hence, the solution we promote is to use *generic benchmarks* that feature a common base for generating various experimental possibilities. The drawback of this approach is that parameter complexity must be mastered, for generic benchmarks to be easily apprehended by users.

In any case, before starting a benchmarking experiment, users' needs must be carefully assessed so that the right benchmark or benchmark class is selected, and test results are meaningful. This sounds like sheer common sense, but many researchers simply select the best known tools, whether they are adapted to their validation experiments or not. For instance, data warehouse papers often refer to TPC-H, while this benchmark's database is not a typical data warehouse, and its workload does not include any OLAP query. *Ad-hoc* and generic benchmarks should be preferred in such situations; and though trust in a benchmark is definitely an issue, relevance should be the prevailing selection criteria. We modestly hope this article will have provided its readers with a fair overview of database benchmarks, and will help them in selecting the right tool for the right job.

Yao, B.B., Ozsu, T., & Khandelwal, N. (2004). XBench Benchmark and Performance Testing of XML DBMSs. 20[th] International Conference on Data Engineering (ICDE 04), Boston, USA. 621-633.

TERMS AND DEFINITION

Database Management System (DMBS): Software set that handles the structuring, storage, maintenance, update and querying of data stored in a database.

Benchmark: A standard program that runs on different systems to provide an accurate measure of their performance.

Synthetic benchmark: A benchmark in which the workload model is artificially generated, as opposed to a real-life workload.

Database benchmark: A benchmark specifically aimed at evaluating the performance of DBMSs or DBMS components.

Database model: In a database benchmark, a database schema and a protocol for instantiating this schema, *i.e.*, generating synthetic data or reusing real-life data.

Workload model: In a database benchmark, a set of predefined read and write operations or operation templates to apply on the benchmark's database, following a predefined protocol.

Performance metrics: Simple or composite metrics aimed at expressing the performance of a system.